\documentclass[journal]{IEEEtran}

\usepackage{cite}
\usepackage{amsmath,amssymb,amsfonts}
\usepackage{graphicx}
\usepackage{textcomp}
\usepackage[T1]{fontenc}
\usepackage{siunitx}
\usepackage{hyperref}
\usepackage{balance}

\begin{document}

\title{High-Velocity Whip-Mode Microresonator in LTOI Unimorph: Measurement Methodology and Large-Signal Characterization}

\author{Tzu-Hsuan~Hsu,
        Zihuan~Liu,
        Harshvardhan~Gupta,
        Ziqian~Yao,
        Wei~Wang,
        Vakhtang~Chulukhadze,
        Jack~Kramer,
        Neal~Hall,
        and~Ruochen~Lu%
\thanks{This work was supported by the DARPA Nimble Ultrafast Microsystems (NIMBUS) program. Any opinions, findings, conclusions, or recommendations expressed in this material are those of the author(s) and do not necessarily reflect the views of the Defense Advanced Research Projects Agency (DARPA).}%
\thanks{T.-H. Hsu, Z. Liu, H. Gupta, Z. Yao, W. Wang, V. Chulukhadze, J. Kramer, N. Hall, and R. Lu are with The University of Texas at Austin, Austin, TX 78712 USA (e-mail: tzuhsuan.hsu@austin.utexas.edu).}%
}


\maketitle

\begin{abstract}
This paper presents the design, characterization, and large-signal measurement methodology of a high-order whip-mode flexural microresonator on a lithium tantalate-on-insulator (LTOI) unimorph platform. A tapered cantilever concentrates kinetic energy at the free tip through a structural velocity amplification effect, with a targeted whip mode at 9.175~MHz exhibiting a measured $Q$ of 691 in air. The results indicate a substantially reduced susceptibility to viscous damping at high modal frequencies. In-air large-signal testing on a separate device confirms tip velocities up to 20~m/s before the reliable measurement range of the laser Doppler vibrometer (LDV) at the tapered tip is exceeded, while the device itself sustains drive levels up to 240~$V_\mathrm{pp}$ before failure. Transitioning to vacuum reveals photothermal-induced static bending of the LTOI cantilever under LDV laser illumination, an effect that prohibits direct velocity measurement for these resonators. Hence, it motivates an indirect extraction methodology to be implemented. In this work, a 3.85 times base-to-tip geometric amplification factor, independently calibrated at low drive, is applied to base velocity measurements to infer tip velocity under large-signal conditions. Using this approach with narrowband chirp excitation, a maximum extracted tip velocity of 58.9~m/s is obtained at 192~$V_\mathrm{pp}$, with spectral analysis of the base velocity placing a conservative lower bound of 36.2~m/s on this estimate. Large-signal failure-mode analysis identifies Pt/Au electrode melting at 210~$V_\mathrm{pp}$ as the current velocity ceiling. These results suggest that geometric amplification in high-order flexural modes offers a viable pathway toward the high proof-mass velocities targeted for next-generation MEMS inertial sensors.

\end{abstract}


\section{Introduction}
\label{sec:introduction}

\IEEEPARstart{S}{ince} the first demonstrations of micromachined vibratory gyroscopes at Draper Laboratory in the late 1980s~\cite{Boxenhorn1988} and the subsequent development of comb-drive tuning fork architectures~\cite{Bernstein1993}, MEMS inertial sensors have undergone rapid improvement across nearly every performance metric. Yet one fundamental relationship has remained underexploited: for Coriolis vibratory gyroscopes~\cite{Hodjat2020,Liu2023}, the scale factor governing sensitivity and dynamic range is directly proportional to proof-mass velocity~\cite{Yazdi1998}. Consequently, substantially increasing proof-mass velocity is among the most direct routes to improving inertial sensor performance. A tenfold increase in velocity directly scales the Coriolis scale factor of a vibratory gyroscope by the same factor. Overall inertial-sensor performance additionally depends on bias stability, quadrature error, mode coupling, and noise sources, and large-amplitude high-velocity operation may itself introduce nonlinearities; velocity is therefore a direct lever on scale factor rather than a universal solution to all performance limits. This relationship has motivated recent community-wide efforts to push MEMS resonators toward fracture-limited velocity regimes. Prior demonstrations of high-velocity MEMS operation have remained limited: silicon resonators have reached velocities on the order of 10~m/s before encountering nonlinear or fracture limits, a 4H-SiC bulk acoustic wave gyroscope achieved particle velocities near 9~m/s~\cite{Liu2024SiC}, and a bulk acoustic wave device exploiting topological interface states in aluminum scandium nitride reached a directly measured out-of-plane velocity of 23~m/s at 82~MHz~\cite{Kaya2025}. Most recently, an aluminum nitride bimorph wedge resonator demonstrated a directly measured tip velocity of 50~m/s at 1.81~MHz in vacuum, driven near the dielectric breakdown limit~\cite{Liu2026}. An earlier comparison of fracture-limited thin-film piezoelectric-on-substrate resonators~\cite{HighVelDiamond2013} reported stored-energy levels from which particle velocities of approximately 30--40~m/s (silicon) and 70--80~m/s (diamond) can be inferred, indicating the potential of high-fracture-strength materials for velocity scaling. The present work extends this line of investigation to a lithium tantalate platform, and addresses the measurement problem that arises once tip velocities exceed what can be acquired directly: whether a calibrated geometric amplification factor can be used to extract tip velocity from a base-point measurement, and how far that extraction can be relied upon.

\begin{figure}[!t]
\centering
\includegraphics[width=\columnwidth]{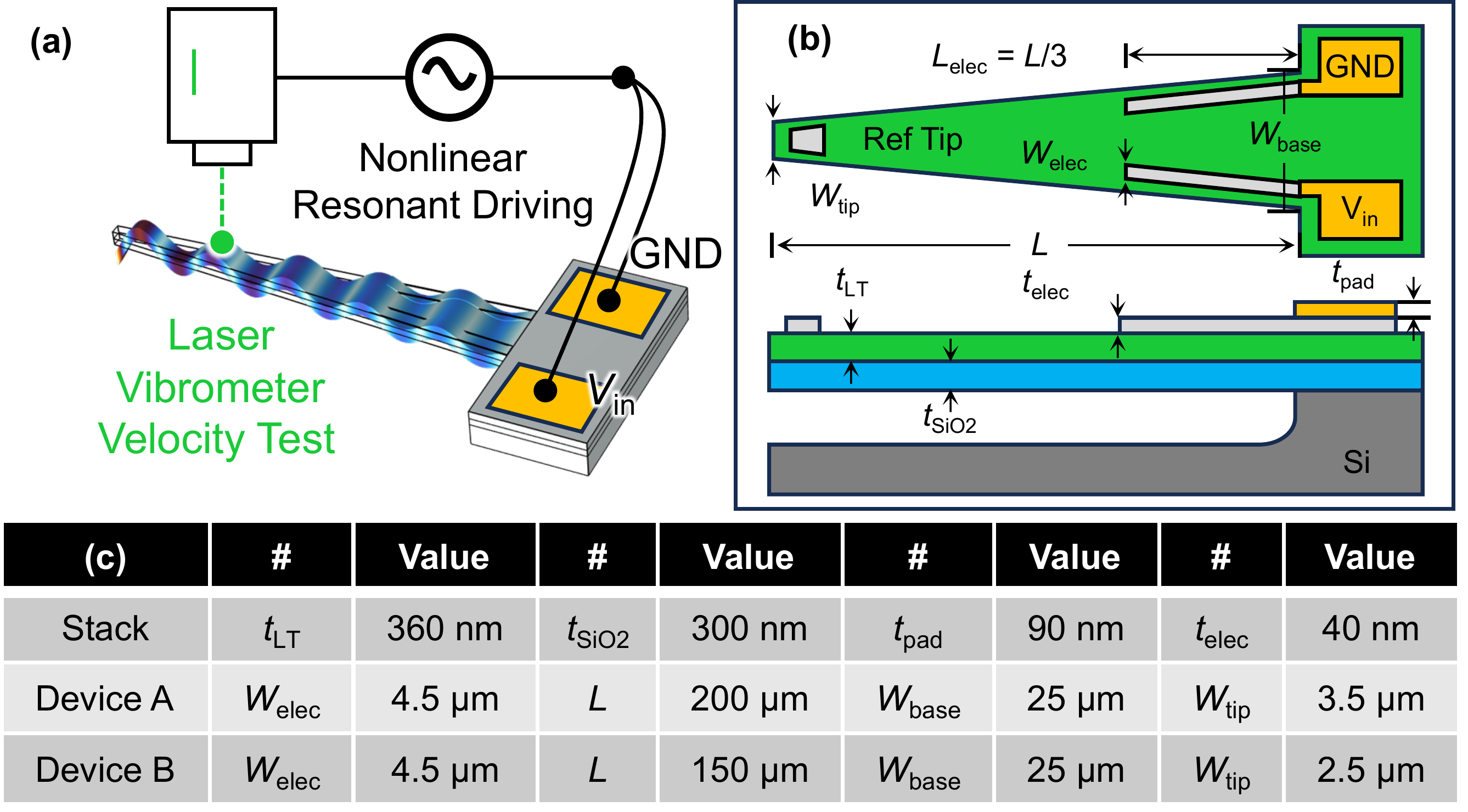}
\caption{(a) Conceptual illustration of a high-order whip-mode flexural resonator and LDV velocity measurement. (b) Top-view layout and side-view stack schematic of the LTOI whip resonator. (c) Key stack thicknesses and geometric parameters for Device~A and Device~B.}
\label{fig:device}
\end{figure}

Despite this motivation, virtually all MEMS resonators to date are deliberately operated in the small-signal linear regime, where displacement is proportional to the drive and device behavior is well understood and predictable. In this regime, performance improvements have historically been pursued through other means: quality factor ($Q$) enhancement~\cite{Asadian2017} and effective-$Q$ engineering through feedback or parametric pumping~\cite{Miller2018Q}, mitigating the impact of thermomechanical noise through improved transduction and readout~\cite{Gabrielson1993}, optimizing readout circuit sensitivity~\cite{Marx2018}, or increasing proof mass~\cite{Yazdi1998} rather than through direct velocity scaling. As a result, practical MEMS resonators remain largely confined to velocities near 5~m/s, leaving the velocity axis of the design space largely unexplored.

\begin{figure}[!t]
\centering
\includegraphics[width=\columnwidth]{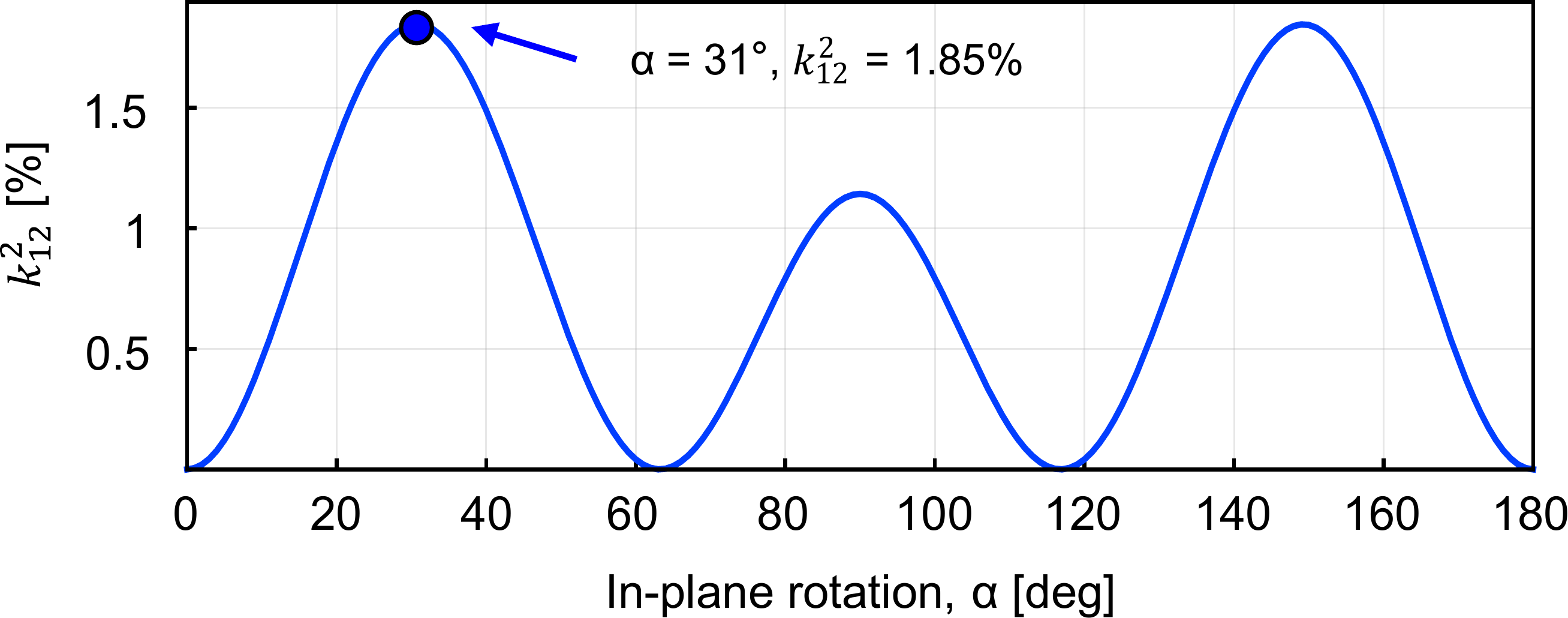}
\caption{Effective lateral-field electromechanical coupling coefficient $k^2_{12}$ as a function of in-plane electrode rotation angle $\alpha$ for 36Y-cut LiTaO\textsubscript{3}, computed via Euler angle transformation of the piezoelectric tensor. The coupling is maximized at $\alpha = 31^\circ$ with $k^2_{12} = 1.85\%$.}
\label{fig:coupling}
\end{figure}

Rather than improving sensitivity through circuit or mass optimization, we directly pursue high resonator velocity as the primary performance lever. Doing so requires addressing several compounding challenges simultaneously. First, low-order flexural modes are highly susceptible to viscous air damping, making large-amplitude operation impractical under ambient conditions and necessitating careful mode selection to reduce susceptibility to air damping~\cite{Tallur2012}. Second, at high drive levels, efficient energy buildup into the resonance requires a drive strategy that can track the nonlinear frequency response. Third, although standard laser Doppler vibrometry (LDV) can enable real-time visualization of out-of-plane resonance modes, large out-of-plane displacements at the free tip of a tapered cantilever can exceed the reliable measurement range of standard LDV, demanding a reliable indirect measurement methodology. Finally, electrode and structural integrity under high-power operation must be characterized to understand the ultimate velocity ceiling. This work addresses each of these challenges in turn.

In this work, we explore these challenges using a high-order flexural mode ``whip'' in a lithium tantalate-on-insulator (LTOI) unimorph cantilever. The high modal order of the targeted whip mode substantially reduces its susceptibility to air damping, while geometric tapering of the cantilever concentrates kinetic energy at the free tip through a structural velocity amplification effect. A narrowband nonlinear chirp drive strategy is employed to efficiently excite the resonance into the large-signal regime. Finally, independent characterization of the base-to-tip velocity amplification ratio provides a calibrated methodology for extracting tip velocity under conditions where direct measurement at the tip is precluded. Using this approach, we estimate an extracted tip velocity of 58.9~m/s for a tethered, chip-scale piezoelectric MEMS resonator under high-nonlinear drive conditions.

\section{Nonlinear Dynamics and Scope}
\label{sec:theory}

Pursuing velocity as the primary performance lever necessarily moves device operation out of the linear regime, and a resonator driven hard enough to reach tens of meters per second will exhibit the nonlinear behavior documented across the MEMS literature. There has been extensive study on the Duffing nonlinearity, which mainly involves an amplitude-dependent frequency shift where the resonance bends with drive level while the energy remains in the driven mode~\cite{Lifshitz2008,Younis2011,Kovacic2011}. This behavior has been modelled and experimentally verified in micromechanical beam resonators~\cite{Shao2008}, and in thin-film piezoelectric resonators driven into the nonlinear regime~\cite{Miller2014NL}.

At the simplest level, the Duffing regime is captured by a single-mode model containing a cubic stiffness term, which reproduces the amplitude-dependent frequency shift and associated hysteresis~\cite{Lifshitz2008,Younis2011,Kovacic2011,Shao2008}. Parametric excitation instead requires a stiffness that varies in time with the drive, and is analyzed by locating the boundaries, in the plane of drive amplitude and drive frequency, beyond which a mode grows rather than decays. This yields the threshold drive level and the frequency bands over which growth occurs~\cite{VanDerAvoort2011}. Energy transfer between distinct modes or the phononic comb formation that can follow from it requires a multimode description in which several modes of the same resonator are coupled through nonlinear terms. Following the formalism of~\cite{Ganesan2018}, a set of nonlinearly coupled mechanical modes driven at $\omega_d$ is described by
\begin{equation}
\label{eq:coupled}
\begin{split}
\ddot{Q}_i = {} & -\omega_i^2 Q_i - 2\zeta_i \omega_i \dot{Q}_i
+ \sum_{\tau_1}^{3}\sum_{\tau_2}^{3} \alpha_{\tau_1 \tau_2} Q_{\tau_1} Q_{\tau_2} \\
& + \sum_{\tau_1}^{3}\sum_{\tau_2}^{3}\sum_{\tau_3}^{3} \beta_{\tau_1 \tau_2 \tau_3} Q_{\tau_1} Q_{\tau_2} Q_{\tau_3}
+ P \cos(\omega_d t),
\end{split}
\end{equation}
where $i = 1, 2, 3$, $Q_i$ is the modal amplitude of mode $i$, $\omega_i$ and $\omega_d$ are the modal and drive frequencies, $\zeta_i$ is the modal damping coefficient, $\alpha$ and $\beta$ are the second- and third-order nonlinear coupling coefficients, and $P$ is the drive amplitude. The expansion is truncated at third order here but may be extended to higher orders. Parametric down-conversion, internal resonance, and coherent comb formation all arise as solutions of this same system. They are distinguished by which coupling coefficients are appreciable, by whether the participating modal frequencies satisfy a resonance condition, and by whether the drive exceeds the corresponding instability threshold~\cite{Qi2020}. Equation~(\ref{eq:coupled}) therefore does not by itself identify which mechanism is operative in a given device. Determining the active mechanism therefore requires dedicated experiments.

Reported behavior in MEMS resonators typically follows a characteristic progression with increasing drive~\cite{Anderson2026,Ganesan2017,Park2019}. In the linear regime, the spectral response remains concentrated at the drive tone. Once the parametric-instability threshold is exceeded, energy can be transferred from the driven mode into lower-frequency modes whose frequencies sum approximately to the drive frequency. With a further increase in drive, nonlinear mixing among the driven and parametrically generated tones can produce phononic frequency combs~\cite{Ganesan2018}. These combs appear as equally spaced spectral sidebands, with their spacing determined by the residual mismatch between the drive frequency and the sum of the parametric-mode frequencies in a cascaded three-wave-mixing process~\cite{Ganesan2017}. Phononic-comb behavior has since been demonstrated in high-$Q$ quartz resonators~\cite{Kubena2020}, reviewed from a broader phenomenological perspective~\cite{Maksymov2022}, and applied to resonance tracking~\cite{GanesanSeshia2019}. Complementary studies have examined the physical and operating conditions governing comb formation, including the effects of drive power and frequency~\cite{Anderson2026}, quality factor~\cite{Qi2020}, and the mechanical or thermal origin of the underlying nonlinearity~\cite{Segovia2013Thermal,LuGong2015}. Other reported routes to multi-line spectra include bifurcation and multimode interactions~\cite{Liu2022,Czaplewski2018} and internal resonance~\cite{Asadi2021,Yu2020,Chen2025,Carvalho2017}, while parametric excitation has also been used to manipulate the temperature coefficient of frequency~\cite{Zheng2024}.

In the reported literature, determination of the nonlinear coefficients is either deferred to dedicated theoretical treatment~\cite{Anderson2026} or constrained by measurements designed for the purpose, in which the participating modes are identified and the instability thresholds are established~\cite{Ganesan2017,Qi2020}. Obtaining these coefficients for the present device would require the same, and lies well beyond what a measurement-focused study of velocity extraction sets out to deliver. The present work instead focuses on experimentally visualizing the behavior of resonators driven to velocities of tens of meters per second.

\begin{figure}[!t]
\centering
\includegraphics[width=\columnwidth]{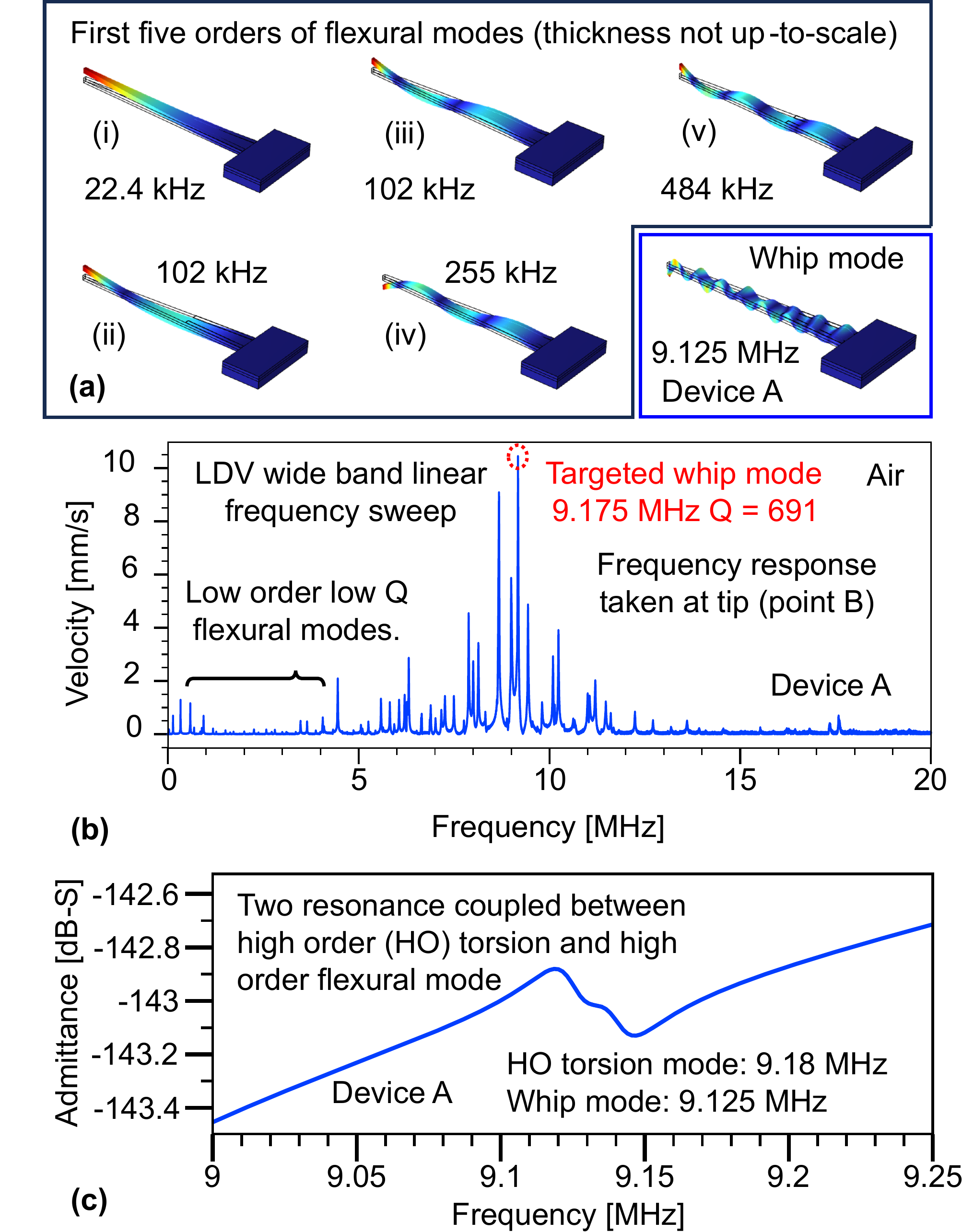}
\caption{(a) Simulated mode shapes of the first five flexural modes and the targeted whip mode that maximizes tip velocity (Device~A). (b) LDV wideband frequency sweep in air measured at the tip, showing the low-$Q$ flexural modes and the dominant whip mode at 9.175~MHz with $Q = 691$. (c) Simulated admittance response near the targeted whip mode, revealing modal coupling between the high-order flexural mode at 9.125~MHz and a high-order torsional mode at 9.18~MHz.}
\label{fig:modes}
\end{figure}

\section{Device Design and Mode Selection}
\label{sec:design}

\subsection{LTOI Unimorph Whip Resonator Design}
\label{subsec:device}

The resonator is built on an LTOI substrate~\cite{Yan2019,Butaud2020} consisting of a 360~nm 36Y-LiTaO\textsubscript{3} (LT) film bonded onto a 300~nm SiO\textsubscript{2} layer on a silicon handle wafer, forming a unimorph structure that enables out-of-plane piezoelectric actuation. Two devices from the same fabrication lot are used in this study: Device~A ($L = \SI{200}{\micro\meter}$, $W_\mathrm{tip} = \SI{3.5}{\micro\meter}$) serves as the primary device for vacuum characterization, while Device~B ($L = \SI{150}{\micro\meter}$, $W_\mathrm{tip} = \SI{2.5}{\micro\meter}$) is used for in-air large-signal testing. Both devices share the same LTOI stack and electrode configuration. The cantilever is tapered in width from a base of \SI{25}{\micro\meter} to the tip over the full cantilever length, with this wedge geometry designed to progressively concentrate strain and kinetic energy toward the free end. The electrode stack consists of a 40~nm platinum layer serving as the primary electrode, capped by a 90~nm gold pad for probing. The top electrode spans the proximal one-third of the cantilever length ($L_\mathrm{elec} = L/3$), positioned in the region of maximum piezoelectric coupling for the targeted mode. The complete device concept and LTOI stack cross-section are shown in Fig.~\ref{fig:device}(a) and~\ref{fig:device}(b), with key geometric and stack parameters for both devices summarized in Fig.~\ref{fig:device}(c). A consolidated summary of both devices, including measured resonance frequencies, quality factors, and measurement environment, is provided in Table~\ref{tab:summary}. A lateral field excitation electrode configuration is used to excite the $e_{12}$ piezoelectric coefficient under an in-plane rotation of $31^\circ$ from the X-axis, corresponding to the maximum lateral-field electromechanical coupling $k^2_{12} = 1.85\%$ for the 36Y-cut LiTaO\textsubscript{3} orientation, as computed via Euler angle transformation of the piezoelectric tensor and shown in Fig.~\ref{fig:coupling}.

The passive SiO$_2$ layer is essential to this transduction scheme. Lateral-field excitation of the $e_{12}$ coefficient produces an in-plane (axial) stress within the piezoelectric film; the passive elastic layer bonded beneath the active film shifts the neutral axis so that this axial stress is converted into a bending moment, generating the out-of-plane flexural motion on which the whip mode relies. Without the passive layer, the axial stress would remain nearly symmetric about the film mid-plane and produce little net out-of-plane bending, and the flexural whip mode could not be efficiently excited. An alternative route to out-of-plane bending that eliminates the passive layer is the periodically-poled piezoelectric film (P3F) approach, in which stacked piezoelectric layers of alternating polarization generate opposite-sign stresses across the film thickness that reinforce the bending moment, as recently demonstrated in bilayer lithium niobate PMUTs~\cite{Chulukhadze2026PMUT,Zhao2025PMUT}. The present device instead achieves out-of-plane transduction using a single, uniformly-oriented piezoelectric film together with a passive elastic layer~\cite{Horsley2021PMUT}, which is simpler to fabricate on the wafer-bonded LTOI substrate.

\begin{table}[!t]
\caption{Summary of Device Parameters and Measured Resonances}
\label{tab:summary}
\centering
\resizebox{\columnwidth}{!}{%
\begin{tabular}{ccccccc}
\hline
Device & $L$ (\si{\micro\meter}) & $W_\mathrm{base}$ (\si{\micro\meter}) & $W_\mathrm{tip}$ (\si{\micro\meter}) & Mode & Freq. & $Q$ (env.) \\
\hline
A & 200 & 25 & 3.5 & Whip & 9.175~MHz & 691 (air) \\
B & 150 & 25 & 2.5 & Whip & 8.03~MHz & 487 (air) \\
B & 150 & 25 & 2.5 & 2nd flexural & 420~kHz & 104 (air) \\
\hline
\end{tabular}%
}
\\[2pt]
{\footnotesize Device~A large-signal characterization was performed in vacuum; all $Q$ values listed are measured in air.}
\end{table}

\subsection{Modal Analysis and Mode Selection}
\label{subsec:modal}

Finite element analysis (FEA) in COMSOL Multiphysics is used to simulate the flexural eigenmodes of the tapered unimorph cantilever. In this analysis, the resonator geometry of Device~A is adopted. The first five flexural mode shapes are shown in Fig.~\ref{fig:modes}(a), with resonant frequencies spanning from 22.4~kHz to 484~kHz. While these lower-order modes exhibit favorable mode shapes with large tip displacement, they are expected to be severely limited by viscous air damping at atmospheric pressure, resulting in low effective $Q$ and limited velocity at practical drive levels.

To effectively evaluate the nonlinear resonator behavior, a whip mode is first identified at 9.125~MHz through simulation. Throughout this work, ``whip mode'' denotes the high-order mode of a given geometry that maximizes tip velocity for a given drive condition. As a result, the corresponding mode order may differ between devices of different geometry. This mode comprises multiple mode-order nodes along the cantilever length, with a concentrated velocity amplitude at the free tip due to a geometric whip effect analogous to the mechanical amplification observed in tapered waveguides. Critically, the high modal frequency substantially reduces its susceptibility to air damping. The frequency response of the simulated admittance in Fig.~\ref{fig:modes}(c) reveals modal coupling between the higher-order flexural whip mode at 9.125~MHz and a nearby torsional mode at 9.18~MHz.

The high electrical impedance of the thin-film LTOI unimorph microresonator renders direct network analyzer-based electrical characterization impractical at the current device scale. Therefore, a 3D broadband linear scan across the entire LTOI resonator was first performed to characterize the device's resonance behavior. The wideband velocity spectrum in Fig.~\ref{fig:modes}(b) was taken at the metal reflector near the tip of Device~A. The results show a series of flexural modes up to 20~MHz. As expected, the low-order flexural modes below 5~MHz exhibit low velocity and reduced $Q$ per unit input due to strong air damping. In contrast, the targeted whip mode at 9.175~MHz clearly emerges as the dominant resonance with the highest velocity amplitude and a measured $Q$ of 691. This mode is selected as the target for all subsequent large-signal characterization. For Device~B, which has a shorter cantilever length and narrower tip (Fig. 1(c)), FEM simulation predicts a corresponding whip mode at 8.03 MHz; this mode is also experimentally confirmed in air and is used exclusively for in-air baseline testing.

\section{Experimental Validation and Results}
\label{sec:results}

\subsection{Large Signal Measurement Setup}
\label{subsec:setup}

To experimentally validate these high-velocity micro resonators, a comprehensive optical characterization workflow is employed to address the high-impedance issue. All large-signal characterization is performed using a scanning laser Doppler vibrometer (Polytec MSA-600)~\cite{Liu2026}, which serves as both the primary velocity-measurement instrument and the time-domain trigger for the measurement. The device is driven by a narrowband periodic chirp burst signal generated internally by the LDV system, amplified by an external power amplifier (Electronics and Innovation A075), and monitored on an oscilloscope prior to application to verify the drive waveform amplitude at each drive level. Under large-signal operation, the resonance frequency of the whip mode shifts appreciably due to Duffing nonlinearity~\cite{Lifshitz2008,Younis2011}. To ensure measurement fidelity across all drive levels, each acquisition is carefully validated by comparing the center frequency of the narrowband chirp drive window against the instantaneous resonance frequency extracted from the measured velocity response. As a result, all large-signal measurements have been confirmed to have the resonance well within the narrow-band driving window. We note that all time-domain traces reported in this work represent the real-time out-of-plane velocity recorded by the LDV during each periodic-chirp burst; the horizontal axis therefore denotes elapsed time (in ms) rather than a frequency axis, and the corresponding frequency-domain content of the response is presented separately via FFT in Fig.~\ref{fig:velocity}(b).

Unless otherwise noted, all velocity values reported in this work denote instantaneous zero-to-peak amplitudes extracted directly from the time-domain LDV records; no RMS, peak-to-peak, or FFT-derived amplitude conventions are used.

Measurement fidelity is verified at every drive level. The MSA-600 decoder supports object velocities up to 150~m/s for vibration frequencies below 300~MHz~\cite{MSA600manual}, an order of magnitude above the largest base velocity reported here. Point~A is located on the flat Pt/Au electrode rails, where the maximum zero-to-peak deflection of $\approx 0.26$~\si{\micro\meter} (15.3~m/s at 9.2~MHz) keeps the local surface tilt well within the collection aperture of the objective and the back-reflected signal level stable. Because the decoded velocity is derived from the frequency of the heterodyne carrier, return-light intensity fluctuations do not bias the extracted velocity while the carrier signal-to-noise ratio remains above the decoder threshold; carrier loss instead produces characteristic broadband spikes. Each acquisition is inspected for such dropout signatures, and affected records are discarded and re-acquired. Finally, the envelope fluctuation observed at large drive (Fig.~\ref{fig:timedomain}) is attributed to the nonlinear spectral redistribution evidenced in Fig.~\ref{fig:velocity}(b), i.e., parametric down-conversion into low-order flexural modes, rather than to a measurement artifact.

\subsection{In-Air Large-Signal Characterization}
\label{subsec:inair}

\begin{figure}[!t]
\centering
\includegraphics[width=\columnwidth]{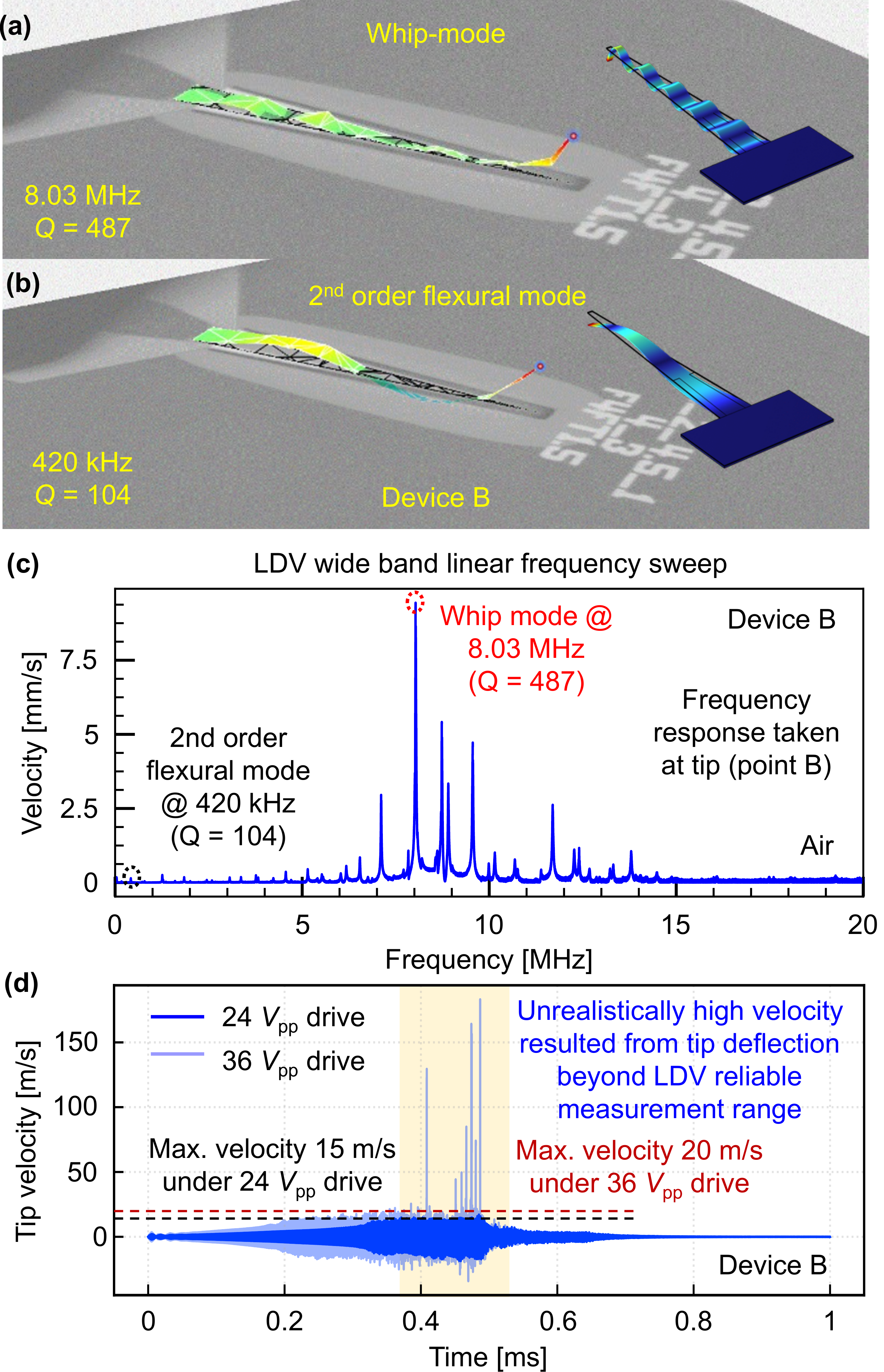}
\caption{In-air characterization of Device~B. (a) LDV-measured and FEM-simulated mode shape of the 8.03~MHz whip mode ($Q = 487$). (b) LDV-measured and FEM-simulated mode shape of the 420~kHz second-order flexural mode ($Q = 104$). (c) LDV wideband frequency sweep in air measured at the tip (Point~B), resolving the 420~kHz second-order flexural mode ($Q = 104$) and the dominant 8.03~MHz whip mode ($Q = 487$), from which the reported quality factors are extracted. (d) Time-domain tip velocity under narrowband chirp excitation at 24~$V_\mathrm{pp}$ and 36~$V_\mathrm{pp}$. At 36~$V_\mathrm{pp}$, unrealistically high velocity spikes (highlighted) indicate that the tip motion has exceeded the reliable LDV measurement range (return-signal degradation and amplitude-dependent spectral impurities).}
\label{fig:inair}
\end{figure}

To establish a baseline for the large-signal behavior of the whip mode prior to vacuum testing, in-air characterization is first performed on Device~B. The 3D LDV scan captures both the low-order flexural modes and the targeted whip mode, as shown in Fig.~\ref{fig:inair}(a) and~\ref{fig:inair}(b). The second-order flexural mode at 420~kHz exhibits $Q = 104$, consistent with dominant viscous air-damping losses at low modal frequencies. In contrast, the whip mode at 8.03~MHz emerges with $Q = 487$, confirming that the high modal frequency substantially reduces---rather than eliminates---the viscous-damping contribution: the whip-mode $Q$ of 487 (Device~B) and 691 (Device~A) in air exceeds that of the low-order flexural mode ($Q \approx 104$) by roughly a factor of five at the same atmospheric pressure. Both quality factors are extracted from the wideband frequency sweep measured at the tip, shown in Fig.~\ref{fig:inair}(c). On Device~A, the corresponding whip mode appears at 9.175~MHz with $Q = 691$ (Fig.~\ref{fig:modes}(b)), with the frequency difference attributable to the different cantilever geometries and mode orders of the two devices.

Under narrowband chirp excitation in air, the tip velocity of Device~B measured directly at Point~B reaches approximately 15~m/s at 24~$V_\mathrm{pp}$ and 20~m/s at 36~$V_\mathrm{pp}$, as shown in Fig.~\ref{fig:inair}(d). At 36~$V_\mathrm{pp}$, the LDV signal at the tip begins to exhibit spurious velocity spikes indicative of the measurement system reaching the limit of its reliable measurement range. The device itself, however, remains mechanically and electrically intact well beyond this drive level: Device~B sustains operation up to 240~$V_\mathrm{pp}$ in air before catastrophic structural failure is observed, indicating substantial headroom between the LDV measurement limit and the device failure threshold. This gap motivates the transition to vacuum, where air-damping losses are eliminated and the full velocity potential of the resonator can be explored---provided the measurement challenges specific to vacuum operation can be addressed.

\subsection{LDV Measurement Constraints Under Large-Signal Vacuum Operation}
\label{subsec:ldv_challenges}

\begin{figure}[!t]
\centering
\includegraphics[width=\columnwidth]{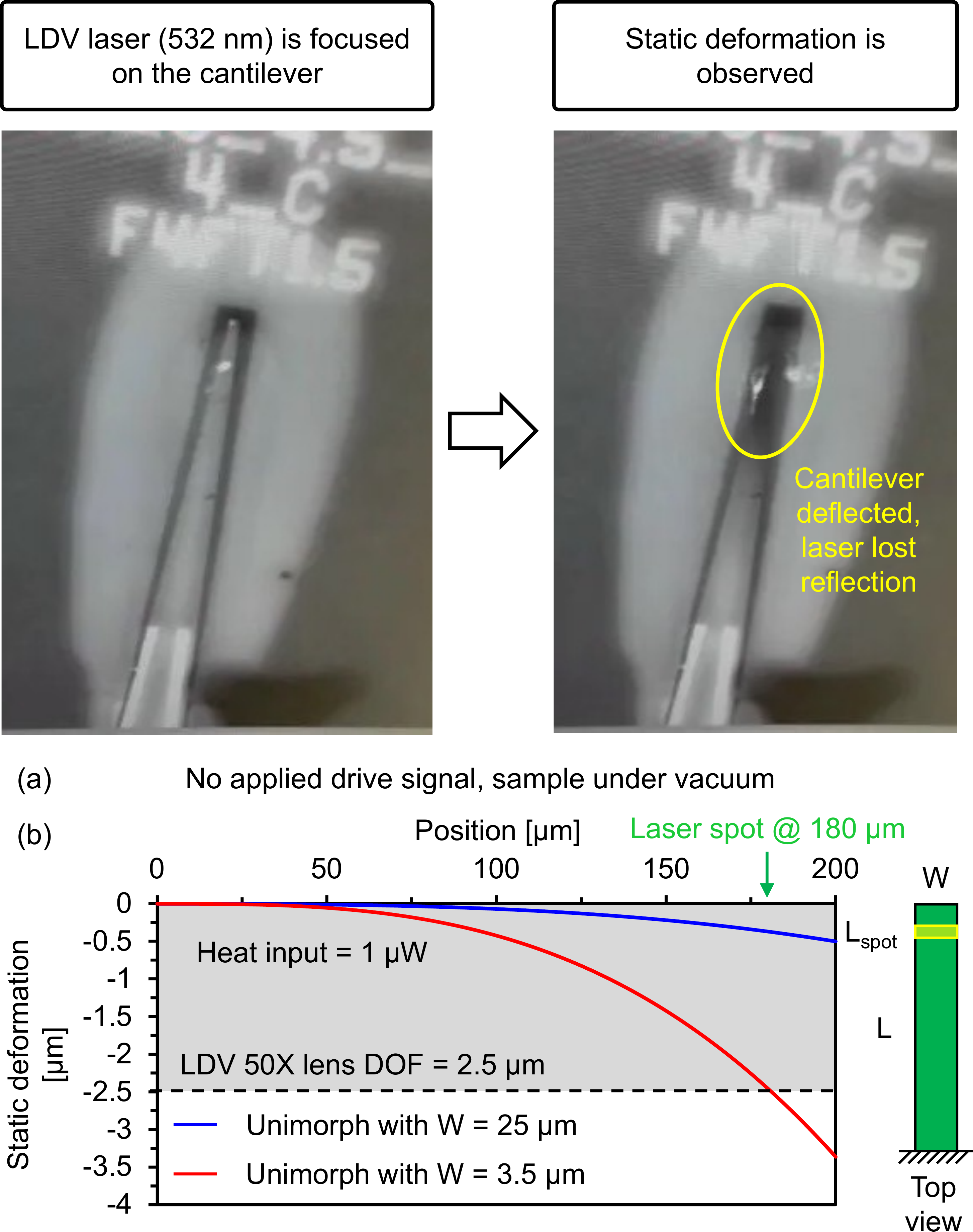}
\caption{Photothermal-induced static deformation of the LTOI cantilever under LDV laser illumination in vacuum with no electrical drive applied. (a) (Left) The 532~nm LDV laser is focused on the cantilever. (Right) Static bending and loss of laser back-reflection are observed, attributed to laser-induced heating of the unimorph stack in the absence of convective cooling. (b) Simulated static deflection profile along the cantilever under \SI{1}{\micro\watt} of absorbed optical power in vacuum, computed for straight beams at the base and tip widths (\SI{25}{\micro\meter} and \SI{3.5}{\micro\meter}) that bracket the tapered geometry ($L = \SI{200}{\micro\meter}$, laser spot at $\sim$$0.9L$). The dashed line marks the \SI{2.5}{\micro\meter} depth of field of the 50$\times$ objective~\cite{MSA600manual}; the shaded band indicates deflections within the focal tolerance.}
\label{fig:photothermal}
\end{figure}

Transitioning from air to vacuum eliminates viscous damping and enables high-amplitude resonant buildup, but introduces two distinct measurement challenges that together preclude direct LDV velocity measurement at the cantilever tip under large-signal drive.

The first challenge concerns the fidelity of tip-point LDV measurement at large amplitude. Although the MSA-600 decoder nominally supports object velocities up to 150~m/s for vibration frequencies below 300~MHz~\cite{MSA600manual}, this specification presumes an amplitude error below 1~dB and a stable back-reflected carrier. At the narrow tapered tip, zero-to-peak deflections approaching $0.4$~\si{\micro\meter} produce large instantaneous surface tilt and periodic defocus that degrade the returned carrier, while large vibration amplitudes introduce spectral impurities in the demodulated signal that scale with the motion amplitude~\cite{MSA600manual}. The resulting decoder dropouts appear as unphysical velocity spikes (Fig.~\ref{fig:inair}(d)) at drive levels well below the device failure threshold, and establish an upper bound on the tip velocity that can be reliably measured.

The second challenge is specific to the vacuum environment. When the 532~nm LDV laser is focused on the LTOI cantilever in vacuum, a pronounced static bending of the cantilever is observed even in the absence of any electrical drive signal, as shown in Fig.~\ref{fig:photothermal}(a). This deformation is not observed under identical laser conditions in air. We attribute this behavior to photothermal heating of the unimorph stack: although stoichiometric LiTaO\textsubscript{3} and SiO\textsubscript{2} are nominally transparent at 532~nm, the Smart~Cut\texttrademark{} ion-implantation process used to fabricate the LTOI substrate is known to introduce lattice damage and sub-bandgap defect states that enable partial optical absorption~\cite{Butaud2020}. Photothermal actuation and laser-driven dynamics of this kind have been studied extensively in MEMS/NEMS cantilevers and related structures~\cite{Aubin2004,Ramos2006,deJong2023,Ganesan2024,Xiao2026}. In air, convective heat transfer dissipates the absorbed energy with negligible temperature rise. In vacuum, the absence of convective cooling leads to a localized temperature increase that, combined with the thermal expansion mismatch between LiTaO\textsubscript{3} and SiO\textsubscript{2}, produces a static unimorph bending. This deformation deflects the cantilever out of the laser focal plane, causing loss of the back-reflection signal at the tip. Reducing the laser intensity to mitigate the heating renders the back-reflected signal insufficient for reliable velocity extraction.

To quantitatively verify the photothermal-induced bending, a stationary thermo-mechanical FEM simulation based on the \SI{360}{\nano\meter} 36Y-LT on \SI{300}{\nano\meter} SiO\textsubscript{2} unimorph beam was performed. Because the optical absorptance of the ion-implanted LTOI film is not independently characterized, in addition to the unmeasurable nature of the exact bending displacement of the beam, an indirect approach of analyzing what absorbed optical power is required to deflect the cantilever beyond the focal tolerance of the LDV with the 50$\times$ objective was used. The clamped end is held at ambient temperature, representing conduction into the silicon substrate, while all remaining surfaces are treated as adiabatic, representing the vacuum environment in which conduction to the anchor is the only heat-loss path. Absorbed power is applied as a surface heat source at the estimated measurement location ($\sim$$0.9L$), with a spot size matching the \SI{1.4}{\micro\meter} nominal beam diameter of the 50$\times$ objective~\cite{MSA600manual}. Since the thermomechanical curvature of a unimorph per unit temperature rise is independent of beam width, the tapered geometry was replaced by straight beams at the base and tip widths (\SI{25}{\micro\meter} and \SI{3.5}{\micro\meter}), which bracket the true structure; width enters only through the thermal resistance to the anchor, and hence through the temperature distribution. As shown in Fig.~\ref{fig:photothermal}(b), the predicted static tip deflection scales linearly with absorbed power, at \SI{0.50}{\micro\meter\per\micro\watt} and \SI{3.37}{\micro\meter\per\micro\watt} for the two bounding widths, the difference reflecting the correspondingly larger temperature rise in the narrower beam. Taking the depth of field (DOF) of the 50$\times$ objective (\SI{2.5}{\micro\meter}, as specified in the instrument documentation~\cite{MSA600manual}) as the criterion for loss of back-reflection, defocusing occurs for an absorbed power of only \SIrange{0.74}{4.97}{\micro\watt}. The instrument is rated Class~3R with a laser output below \SI{5}{\milli\watt} at \SI{532}{\nano\meter}~\cite{MSA600manual}; even for a conservatively assumed delivered power of \SI{0.5}{\milli\watt}, the required absorptance remains at or below approximately 1\%, and falls to below 0.1\% at the full rated power. Absorption at this level is entirely consistent with the lattice damage and sub-bandgap defect states introduced by the Smart~Cut\texttrademark{} ion-implantation process~\cite{Butaud2020}. This analysis supports photothermal heating as a physically plausible explanation for the deformation observed in Fig.~\ref{fig:photothermal}(a).

The two constraints, including return-signal degradation at the tip under large-amplitude drive and photothermal-induced signal degradation in vacuum motivate the indirect velocity-extraction methodology described in the following section. By measuring at a point near the clamped base (Point~A), where displacements remain within the reliable measurement range, and the metallic electrode pad provides robust laser back-reflection unaffected by photothermal bending, a reliable base velocity measurement can be obtained and scaled to the tip via the geometric amplification factor.

\subsection{Geometric Amplification as a Velocity Extraction Methodology}
\label{subsec:amplification}

\begin{figure}[!t]
\centering
\includegraphics[width=\columnwidth]{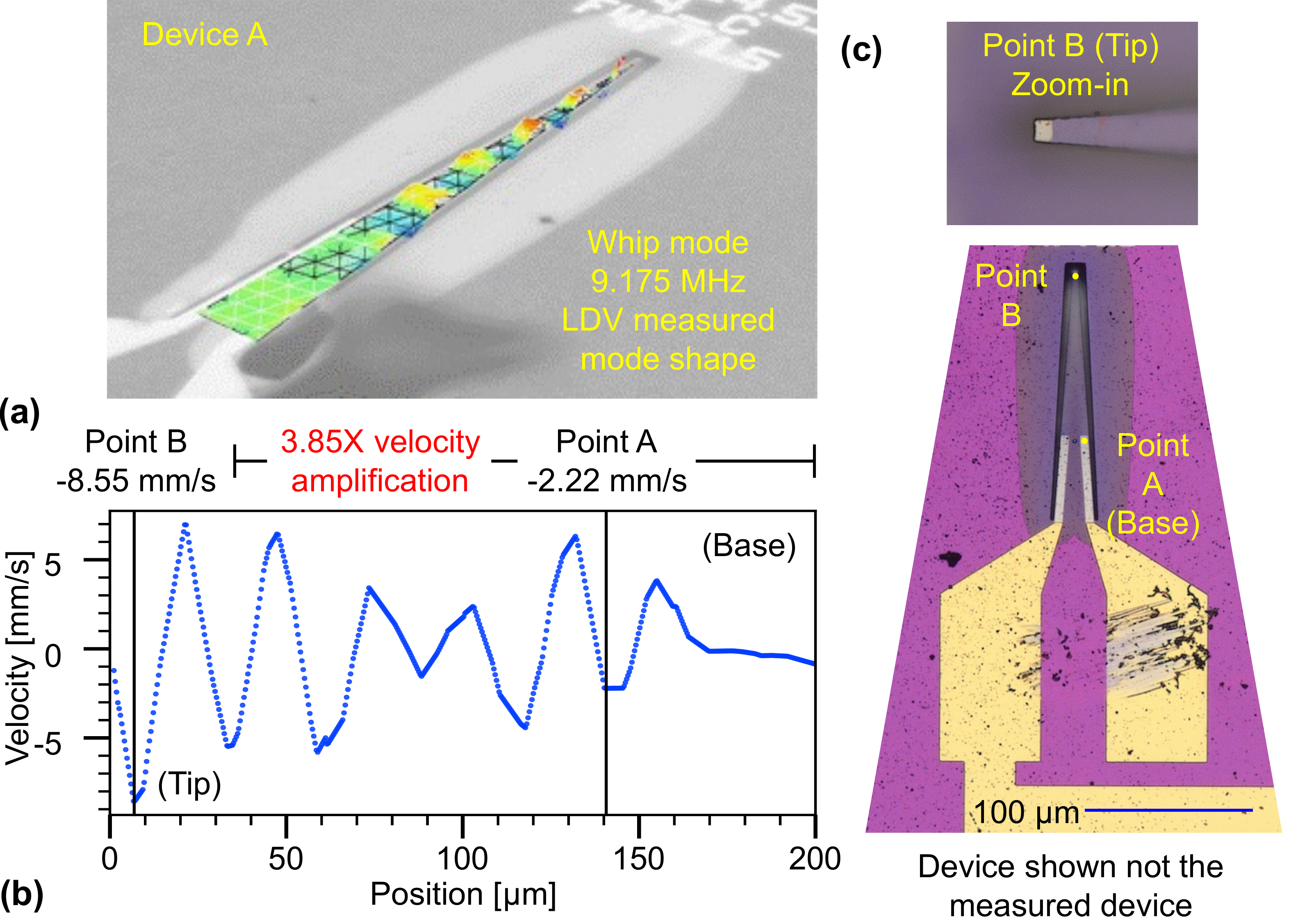}
\caption{Geometric amplification characterization of Device~A. (a) LDV-measured mode shape at the whip resonance 9.175~MHz. (b) Measured velocity profile along the cantilever length showing 3.85 times base-to-tip amplification (Point~A to Point~B). (c) Optical microscope image of the whip resonator showing the LDV measurement locations.}
\label{fig:amplification}
\end{figure}

At low drive amplitudes where both the tip and base regions are within the LDV reliable measurement range, the velocity profile along the full cantilever length of Device~A is measured point-by-point at the whip resonance, as shown in Figs.~\ref{fig:amplification}(a) and~\ref{fig:amplification}(b). Point~A, located near the proximal electrode region, exhibits minimal displacement with reliable LDV back-reflection due to the flat, well-defined surface of the electrode rails. Point~B, near the free tip, exhibits the maximum displacement but poses challenges for direct measurement at high drive amplitudes due to the measurement-range and photothermal constraints described in previous sections. The ratio of the peak velocity amplitudes at these two locations yields a geometric amplification factor of 3.85 times (Point~B: $-8.55$~mm/s; Point~A: $-2.22$~mm/s). This ratio is an empirical property of the measured operating mode shape and electrode configuration. It therefore provides the scaling factor used to infer tip velocity from base measurements at high drive levels in vacuum; its applicability under large-signal drive is examined in Section~\ref{sec:discussion}. We note that the neighboring resonances visible in the wideband spectrum (Fig.~\ref{fig:modes}(b)) are not excited under the narrowband chirp used here, so only a single resonance is measured in every large-signal acquisition; the Duffing nonlinearity shifts the whip-mode frequency within the drive window but does not bring these neighbors into the excitation band. The flexural--torsional coupling indicated in Fig.~\ref{fig:modes}(c) manifests as a single hybrid mode whose combined motion is resolved directly by the 3D LDV scan; this coupled character is also why we refer to the operating mode as a whip mode rather than a purely high-order flexural mode. Because the 3.85 times amplification factor is extracted from the measured velocity profile of this actual operating mode, any such coupling is inherently embedded in the measured ratio rather than biasing it.

For the same reason, the measured profile in Fig.~\ref{fig:amplification}(b) is not expected to reproduce the purely flexural eigenmodes of Fig.~\ref{fig:modes}(a). The local peak velocity does not fall off smoothly from the tip toward the base: it drops to a minimum near mid-span and then grows again. A single flexural mode cannot produce this, whereas two modes of slightly different wavelength present at once can, consistent with the flexural--torsional coupling shown in Fig.~\ref{fig:modes}(c), and with the coupled flexural--torsional dynamics reported for beams of low torsional stiffness~\cite{Carvalho2017}. Because the amplification factor is extracted from the measured profile of the actual operating mode rather than from a simulated flexural mode shape, this difference does not propagate into the extraction.

\subsection{Large-Signal Vacuum Characterization}
\label{subsec:vacuum}

\begin{figure}[!t]
\centering
\includegraphics[width=\columnwidth]{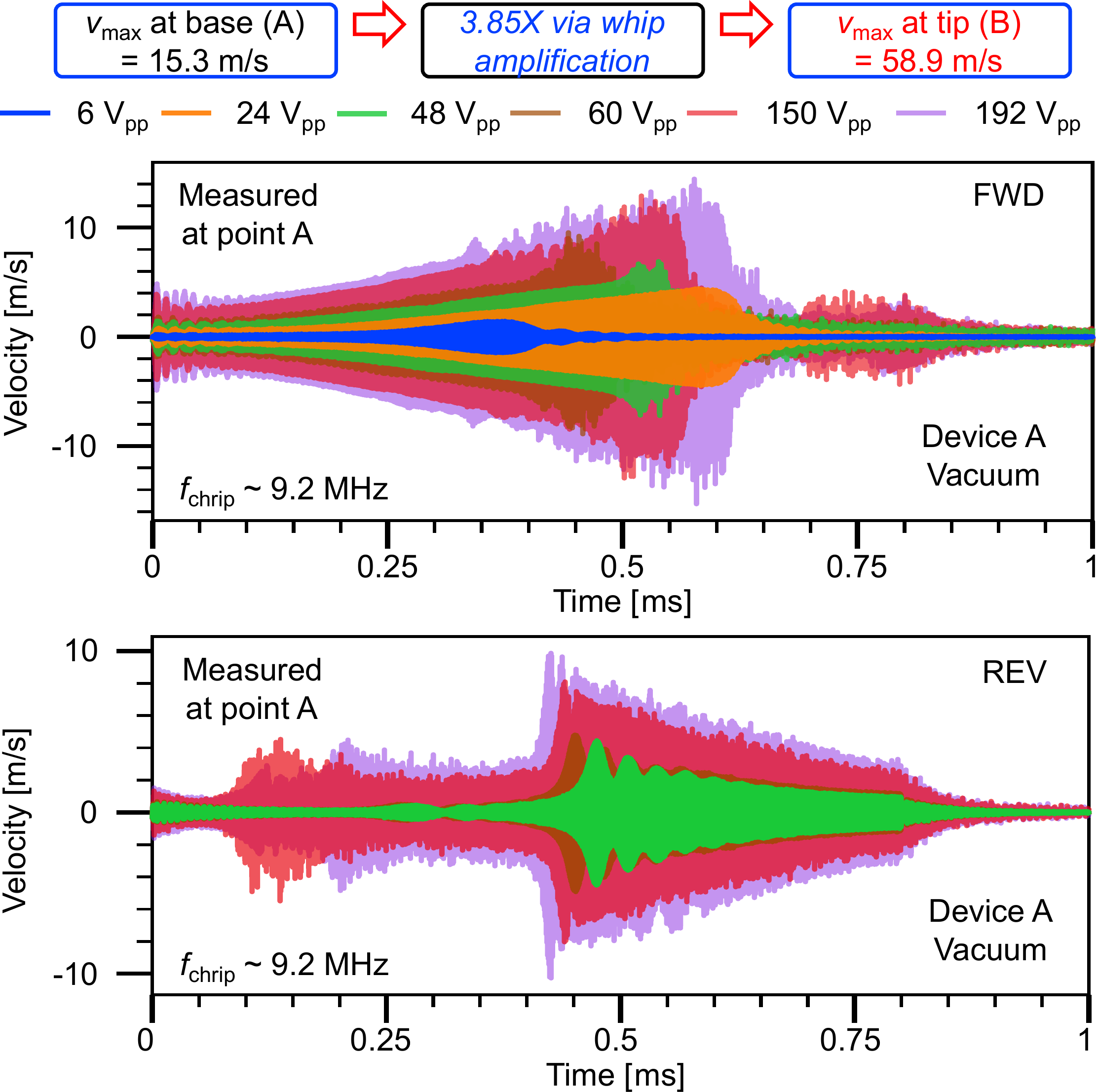}
\caption{LDV-measured time-domain base velocity at Point~A of Device~A in vacuum under forward (FWD) and reverse (REV) narrowband chirp excitation (centered near 9.2~MHz) for progressively increased drive amplitudes from 6~$V_\mathrm{pp}$ to 192~$V_\mathrm{pp}$.}
\label{fig:timedomain}
\end{figure}

\begin{figure}[!t]
\centering
\includegraphics[width=\columnwidth]{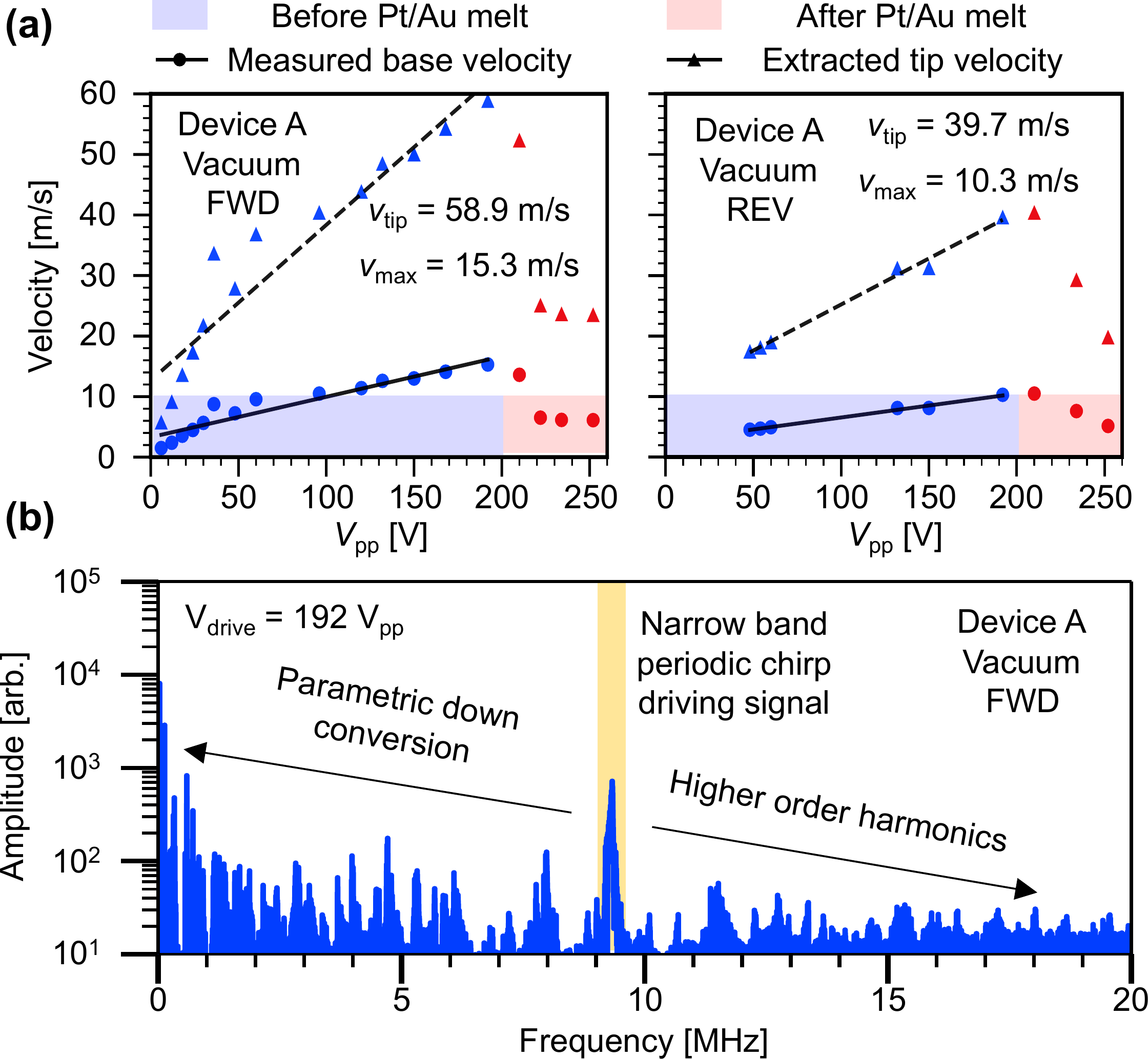}
\caption{(a) Measured base velocity and extracted tip velocity versus drive voltage for Device~A in vacuum under FWD and REV chirp excitation. Data points before Pt/Au electrode melting (blue region) and after melting (red region) are distinguished. (b) FFT of the base velocity time-domain signal at 192~$V_\mathrm{pp}$ under FWD excitation, showing nonlinear spectral redistribution including parametric down-conversion to low-frequency components and higher-order harmonics. Uncertainty bounds for the extracted tip velocities in (a) are listed in Table~\ref{tab:uncertainty} and discussed in Section~\ref{sec:discussion}.}
\label{fig:velocity}
\end{figure}

Large-signal testing is then conducted in vacuum on Device~A to eliminate air-damping losses and enable high-amplitude resonant buildup. The device is excited by a narrowband periodic chirp burst signal centered near the whip-mode frequency ($\sim$9.2~MHz). Both forward (FWD) and reverse (REV) chirp sweeps are performed to characterize hysteresis and sweep-direction dependence of the nonlinear response.

Fig.~\ref{fig:timedomain} shows the time-domain base velocity measured at Point~A under forward and reverse chirp excitation for six drive amplitudes ranging from 6~$V_\mathrm{pp}$ to 192~$V_\mathrm{pp}$. In both sweep directions, the velocity envelope grows progressively with increasing drive level, with peak base velocity reaching 15.3~m/s under FWD excitation at 192~$V_\mathrm{pp}$. Applying the independently characterized 3.85 times geometric amplification factor, the maximum extracted tip velocity reaches 58.9~m/s for the FWD sweep. We note that this value is inferred from the base measurement via the independently characterized geometric amplification factor, and assumes preservation of the mode shape under large-signal drive conditions; this assumption is examined quantitatively in Section~\ref{sec:discussion}.

The corresponding extracted tip velocity scaling with drive amplitude is shown in Fig.~\ref{fig:velocity}(a) for both FWD and REV conditions, alongside the directly measured base velocity. A clear asymmetry between FWD and REV sweeps is observed, with FWD excitation consistently yielding higher peak velocities. This is a signature of the Duffing nonlinearity, which introduces a sweep-direction dependence in the energy buildup. The velocity scaling with $V_\mathrm{pp}$ also shows a clear change in slope associated with electrode degradation, as discussed in the next section.

To further explore the nonlinearity characteristics of such devices, FFT analysis of the base velocity time-domain signal at 192~$V_\mathrm{pp}$, shown in Fig.~\ref{fig:velocity}(b), reveals strong nonlinear spectral redistribution under large-signal excitation. Energy injected near the 9.2~MHz whip mode is redistributed to a cluster of low-frequency components and higher-order harmonics spanning the measurement bandwidth. This redistribution is consistent with parametric down-conversion and multimode coupling, although the present measurement does not uniquely identify the underlying mechanism. Similar spectral redistribution has also been reported in large-signal MEMS resonator studies~\cite{Liu2022,Park2019}. Parametric conversion, multimode interaction, and coherent comb formation can each produce comparable signatures in a single-point spectrum, and the base-velocity acquisition used here cannot discriminate among them. The transient chirp excitation also differs from the conditions under which coherent combs have been established~\cite{Ganesan2017,Anderson2026}. Hence, the spectrum is used only to establish the methodological consequence that a fraction of the injected energy lies outside the whip band and therefore bounds the validity of the velocity extraction.

\subsection{Large Signal Failure Mode Analysis}
\label{subsec:failure}

\begin{figure}[!t]
\centering
\includegraphics[width=\columnwidth]{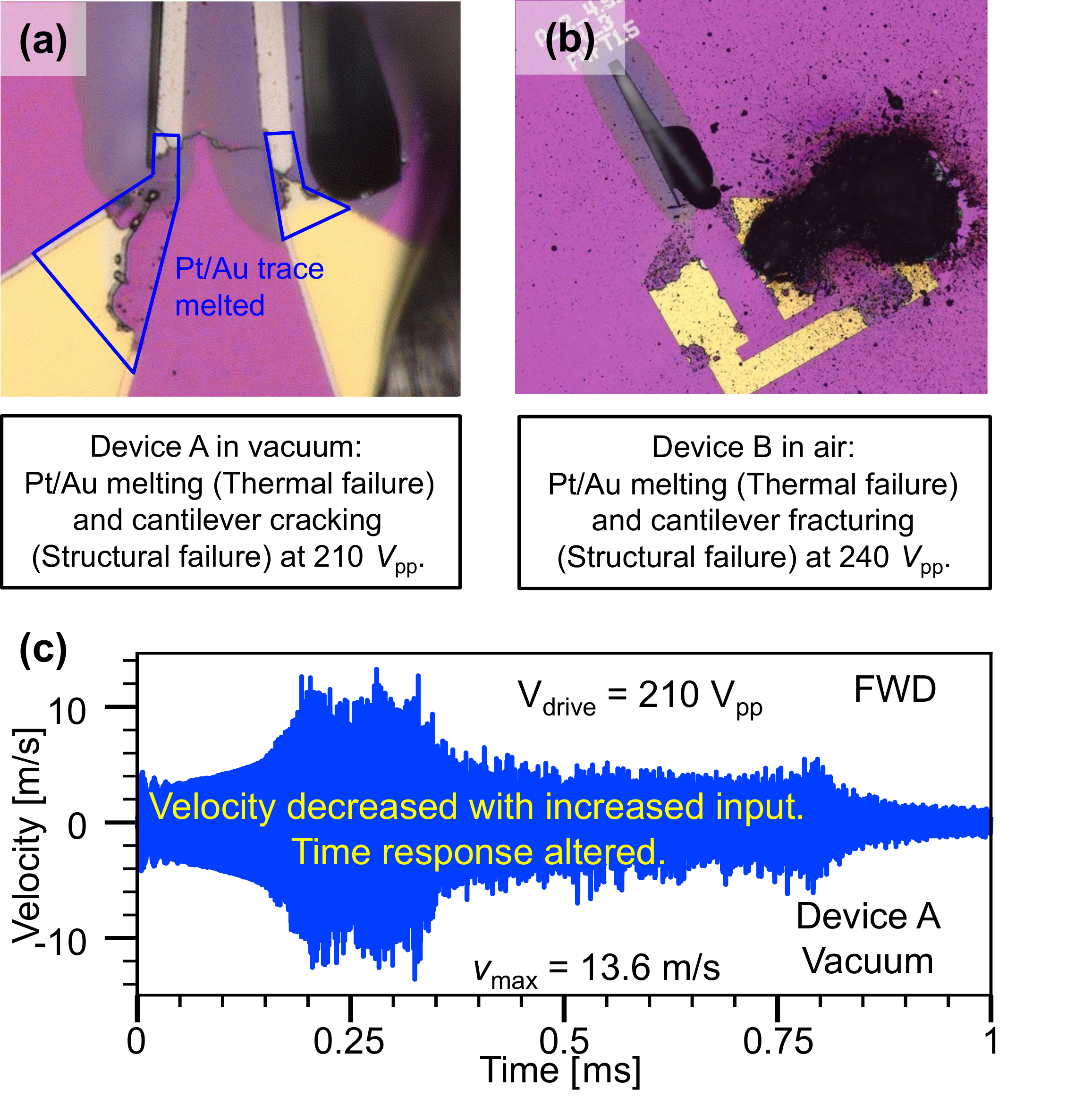}
\caption{Large-signal failure mode analysis. (a) Post-test optical image of Device~A after 210~$V_\mathrm{pp}$ drive in vacuum, showing Pt/Au electrode trace melting and cantilever cracking. (b) Post-test optical image of Device~B after 240~$V_\mathrm{pp}$ drive in air, showing electrode melting and catastrophic cantilever fracture. (c) Time-domain velocity response of Device~A at 210~$V_\mathrm{pp}$ in vacuum, showing reduced peak velocity (13.6~m/s vs.\ 15.3~m/s at 192~$V_\mathrm{pp}$) and qualitatively altered waveform envelope consistent with sudden electrode degradation.}
\label{fig:failure}
\end{figure}

To characterize the device's large-signal input ceiling, the drive amplitude of Device~A is progressively increased beyond 192~$V_\mathrm{pp}$. Fig.~\ref{fig:failure}(c) shows the time-domain velocity response at 210~$V_\mathrm{pp}$, immediately after which visible alteration of the electrode traces is observed. At this drive level, the measured base velocity drops from 15.3~m/s to 13.6~m/s, and the time-domain waveform exhibits a qualitatively altered envelope shape. Post-test inspection confirms melting of the Pt/Au electrode traces and cracking of the cantilever (Fig.~\ref{fig:failure}(a)), identifying ohmic heating under high current density as the dominant failure mechanism in vacuum. For comparison, Device~B sustains operation in air up to 240~$V_\mathrm{pp}$ before exhibiting both electrode melting and catastrophic cantilever fracture (Fig.~\ref{fig:failure}(b)), consistent with the additional convective cooling available in air, delaying the onset of thermal failure. This thermal failure mode defines the current velocity ceiling. Increasing the Au capping layer thickness or widening the electrode traces to reduce current density offers a straightforward path to raising the thermal failure threshold in future device iterations.

\section{Discussion and Limitations}
\label{sec:discussion}


The tip velocities reported in Fig.~\ref{fig:velocity}(a) are estimates. The calibrated amplification factor is applied to the measured base velocity under two assumptions: that the factor calibrated at low drive remains applicable at large drive, and that the full-bandwidth base velocity scales to the tip with that same factor. Both are examined below.

Whether the whip mode remains the dominant response at large drive is assessed directly from the base-velocity records of Fig.~\ref{fig:timedomain}. Each acquisition comprises $4 \times 10^{5}$ samples at a \SI{20}{\nano\second} sampling interval. The whip-mode line is located as the spectral maximum between 8.6 and \SI{9.9}{\mega\hertz}, a window wide enough to accommodate the Duffing shift with drive level, and the modal energy fraction is evaluated as the fraction of total spectral power falling within $\pm$\,\SI{0.5}{\mega\hertz} of that line. The whip band holds more than $83\%$ of the base-velocity energy at every drive level, reaching $92.6\%$ at the highest reported drive of 192~$V_\mathrm{pp}$. The targeted mode therefore continues to dominate the response throughout the reported range, although geometric nonlinearities can in principle modify mode shapes at large amplitude~\cite{Givois2021} and direct confirmation at the tip is precluded by the reliable measurement range.

The second assumption is one-sided in its consequences. It should be noted that the energy fraction above and the bound below are different statistics: the former integrates over the full record, whereas the reported velocity is an instantaneous peak, to which a low-energy but impulsive component can contribute disproportionately. The reported values apply the amplification factor to the full-bandwidth time-domain peak, which at large drive also contains the parametrically generated content described in Section~\ref{subsec:vacuum}. Whether that content adds to or subtracts from the tip motion cannot be determined here, since both its base-to-tip ratio and its phase relative to the whip mode are unknown. The most conservative bound available is therefore obtained by retaining only the whip-mode band and discarding everything else, giving 36.2~m/s against the reported 58.9~m/s at 192~$V_\mathrm{pp}$. Table~\ref{tab:uncertainty} lists this bound alongside the reported values across the drive range. The reported value is retained as the estimate, since discarding the remaining content assumes that it contributes nothing at all, which is equally unlikely; the bound indicates how much of the estimate rests on the assumption rather than replacing it. The amplification factor itself is extracted from the spatially resolved 3D LDV velocity scan of Fig.~\ref{fig:amplification}(b) rather than from two isolated point acquisitions, and both calibration points lie in regions where the measured velocity varies slowly with position, which limits the sensitivity of the ratio to laser placement.

Finally, while the extracted tip velocity of 58.9~m/s is comparable to the 50~m/s directly measured in a companion AlN bimorph wedge resonator~\cite{Liu2026}, the two results are complementary: the AlN device benefits from direct tip-velocity LDV measurement but is limited by dielectric breakdown near 400~V, whereas the LTOI device circumvents the LDV measurement-range limitation through geometric amplification but is instead limited by electrode thermal failure at 210~$V_\mathrm{pp}$. The LTOI platform further offers a distinct material basis in single-crystal piezoelectric coupling ($e_{12}$), motivating continued exploration of lithium-tantalate-based resonators for high-velocity MEMS applications.

\begin{table}[!t]
\caption{Extracted Tip Velocity and Conservative Lower Bound}
\label{tab:uncertainty}
\centering
\resizebox{\columnwidth}{!}{%
\begin{tabular}{cccccc}
\hline
Drive & \multicolumn{2}{c}{Base velocity (m/s)} & Whip-band & \multicolumn{2}{c}{Tip velocity (m/s)} \\
\cline{2-3}\cline{5-6}
($V_\mathrm{pp}$) & full band & whip band & energy (\%) & reported & lower bound \\
\hline
  6 &  1.5 &  1.4 & 93.5 &  5.8 &  5.4 \\
 24 &  4.5 &  4.1 & 99.5 & 17.3 & 15.8 \\
 48 &  7.2 &  5.1 & 90.3 & 27.7 & 19.6 \\
 60 &  9.6 &  6.0 & 83.8 & 37.0 & 23.1 \\
150 & 13.0 &  8.5 & 93.7 & 50.1 & 32.7 \\
192 & 15.3 &  9.4 & 92.6 & 58.9 & 36.2 \\
\hline
\end{tabular}%
}
\\[2pt]
{\footnotesize Tip velocities are 3.85 times the corresponding base velocity. The whip band is $\pm$\,\SI{0.5}{\mega\hertz} about the resonance. Reported values use the full-bandwidth time-domain peak; the lower bound retains the whip band only. FWD excitation, Device~A in vacuum.}
\end{table}


Resolving the nonlinear-mechanism question calls for a dedicated study including threshold measurements across a finely sampled drive range, spatially resolved acquisition to identify which modes receive the down-converted energy, and phase-coherence analysis of the generated lines. The flexural--torsional coupling evident in Fig.~\ref{fig:modes}(c) makes this device family a reasonable candidate for such a study. On the measurement side, future work should validate the base-to-tip amplification factor at full drive using LDV with an extended range, investigate direct electrical readout using a high-input-impedance buffer or transimpedance amplifier, and quantify long-term stability and aging under sustained excitation. Furthermore, as the tapered device has a small effective modal mass, translating its high velocity into useful momentum and gyroscope scale factor will additionally require proof-mass co-design.

\section{Conclusions}
\label{sec:conclusions}

This work has presented the design, large-signal characterization, and measurement methodology of a high-order whip-mode flexural microresonator based on an LTOI unimorph platform. In-air testing confirms that the high modal frequency of the whip mode substantially reduces its susceptibility to air damping, with $Q = 691$. Yet, two main challenges exist during the velocity characterization of these devices. First, large vibratory amplitudes under high-power drive often exceed the reliable LDV measurement range at the tapered tip well before device failure. In addition, photothermal-induced bending of the unimorph cantilever unique to vacuum measurement under LDV laser illumination introduces an additional measurement constraint. Hence, this motivates the development of an indirect velocity extraction approach based on a 3.85 times geometric amplification factor calibrated at the cantilever base. Applying this methodology under narrowband chirp excitation yields an extracted tip velocity of 58.9~m/s at 192~$V_\mathrm{pp}$, with a conservative lower bound of 36.2~m/s obtained by retaining only the whip-mode spectral content of the base velocity. Thermal failure of the Pt/Au electrodes at 210~$V_\mathrm{pp}$ identifies the current performance ceiling, with electrode material and geometry optimization offering a clear path toward higher drive levels. The measurement framework presented here, combining geometric amplification and base-point LDV acquisition, may be broadly applicable to high-velocity MEMS resonators where direct tip measurement is impractical, and the velocity results support continued exploration of high-order flexural modes as a scaling lever for MEMS inertial sensor performance.

\section*{Acknowledgments}
The authors would like to thank Dr.~Sunil Bhave for helpful discussions and Soitec for providing the LTOI substrates.

\bibliographystyle{IEEEtran}

\balance

\end{document}